\newcommand{\be}{\begin{equation} }
\newcommand{\ee}{\end{equation} }
\newcommand{\bea}{\begin{eqnarray} }
\newcommand{\eea}{\end{eqnarray} }

\documentclass[aps,prl,reprint,showpacs,longbibliography,superscriptaddress, citeonline]{revtex4-2}
\usepackage{amsmath,amssymb,bm}
\usepackage{graphicx}
\usepackage{xspace}
\usepackage{latexsym}
\usepackage{color}
\usepackage{hyperref}
\usepackage{orcidlink}

\newcommand{\DTNX} {Ni(Cl$_{1-x}$Br$_x$)$_2$$\cdot$4SC(NH$_2$)$_2$\xspace}

\newcommand{\SrZn}{SrZnVO(PO$_4$)$_2$\xspace}

\newcommand{\BaCd}{BaCdVO(PO$_4$)$_2$\xspace}

\newcommand{\PbV}{Pb$_2$VO(PO$_4$)$_2$\xspace}

\begin{document}
\title{Narrowly avoided spin-nematic phase in BaCdVO(PO$_4$)$_2$: NMR evidence}

\author{K.~M.~Ranjith\,\orcidlink{0000-0001-8681-2461}}
\affiliation{
Laboratoire National des Champs Magn\'{e}tiques Intenses, LNCMI-CNRS (UPR3228), EMFL, Universit\'{e} \\
Grenoble Alpes, UPS and INSA Toulouse, Bo\^{\i}te Postale 166, 38042 Grenoble Cedex 9, France}
\affiliation{Present address: IFW Dresden, Helmholtzstra{\ss}e 20, 01069 Dresden, Germany}	
\author{K.~Yu.~Povarov\,\orcidlink{0000-0002-0026-743X}}
\affiliation{Laboratory for Solid State Physics, ETH Z\"{u}rich, 8093 Z\"{u}rich, Switzerland}
\affiliation{Present address: Dresden High Magnetic Field Laboratory (HLD-EMFL) and W\"urzburg-Dresden \\
Cluster of Excellence ct.qmat, Helmholtz-Zentrum Dresden-Rossendorf, 01328 Dresden, Germany}	
\author{Z. Yan}
\affiliation{Laboratory for Solid State Physics, ETH Z\"{u}rich, 8093 Z\"{u}rich, Switzerland}
\author{A.~Zheludev\,\orcidlink{0000-0002-4972-0709}}
\homepage{http://www.neutron.ethz.ch/}
\affiliation{Laboratory for Solid State Physics, ETH Z\"{u}rich, 8093 Z\"{u}rich, Switzerland}
\author{M.\,Horvati{\'c}\,\orcidlink{0000-0001-7161-0488}}
\email{mladen.horvatic@lncmi.cnrs.fr}
\affiliation{
Laboratoire National des Champs Magn\'{e}tiques Intenses, LNCMI-CNRS (UPR3228), EMFL, Universit\'{e} \\
Grenoble Alpes, UPS and INSA Toulouse, Bo\^{\i}te Postale 166, 38042 Grenoble Cedex 9, France}

\begin{abstract}
We present a $^{31}$P nuclear magnetic resonance (NMR) investigation of BaCdVO(PO$_4$)$_2$ focusing on the nearly saturated regime between $\mu_0H_{c1} = 4.05$~T and $\mu_0H_{c2} = 6.5$~T, which used to be considered a promising candidate for a spin-nematic phase. NMR spectra establish the absence of any dipolar order there, whereas the weak field dependence of the magnetization above $H_{c1}$ is accounted for by Dzyaloshinskii-Moriya interaction terms. The low-energy spin dynamics (fluctuations), measured by the nuclear spin-lattice relaxation rate $T_1^{-1}$, confirms the continuity of this phase and the absence of any low-temperature phase transition. Unexpectedly, the spin dynamics above $H_{c1}$ is largely dominated by  two-magnon processes, which is expected \emph{above} the saturation field of a spin-nematic phase, but not inside. This shows that BaCdVO(PO$_4$)$_2$ is indeed close to a spin-nematic instability; however, this phase is \emph{not} stabilized. We thus confirm recent theoretical predictions that the spin-nematic phase can be stabilized, at most, in an extremely narrow field range close to saturation or is rather narrowly avoided [Jiang \textit{et al.}, \href{https://doi.org/10.1103/PhysRevLett.130.116701}{Phys. Rev. Lett. \textbf{130}, 116701 (2023)}].
\end{abstract}

\date{\today}
\maketitle

\section{Introduction}
\vspace{-0.2cm}
One good place to look for exotic quantum phases in magnetic insulators is in frustrated antiferromagnets close to a classical ferromagnetic (FM) instability \cite{Ueda2013,Starykh2014}. On very general arguments, in the applied magnetic field these systems either enter saturation in a (weakly) discontinuous transition or exhibit some kind of purely quantum ``presaturation phase.'' One of the most famous examples is the $S=1/2$ Heisenberg square-lattice model with FM nearest-neighbor (NN) coupling $J_1$ and frustrating antiferromagnetic (AFM) next-nearest-neighbor (NNN) interaction $J_2$. Near saturation, it has been predicted to support the so-called spin-nematic state \cite{Shannon2006,Ueda2013,Ueda2015} for a wide range of frustration ratios $J_2/J_1$. The simple physical picture is condensation of bound magnon pairs stabilized by the FM bonds. The energy of such pairs is lower than that of two free magnons. Thus, upon lowering the field in the fully polarized state, they condense before the conventional one-magnon Bose-Einstein Condensation, responsible for the N\'{e}el (dipolar) order, can take place.

To date, this type of multipolar ordered phase has remained as experimentally elusive as quantum spin liquids. The materials that most closely correspond to this model are the quasi-two-dimensional (quasi-2D) layered vanadyl phosphates with the general formula  $AA^\prime$VO(PO$_4$)$_2$  ($A$, $A^\prime$ = Ba, Cd, Pb, Sr, Zn) \cite{Nath2008,TsirlinSchmidt2009,TsirlinRosner2009}. In at least two compounds, \PbV and \SrZn, rather unusual narrow presaturation phases have been discovered, but have been shown to represent some kind of complex dipolar-ordered, rather than nematic, states \cite{Landolt2022,Ranjith2022,Landolt2021,Landolt2020}.

Of all these compounds the most frustrated and the most promising spin-nematic candidate is \BaCd \cite{TsirlinSchmidt2009,TsirlinRosner2009}. Indeed, recent studies provided  thermodynamic and neutron diffraction evidence that this material may have a novel exotic quantum phase in applied fields above $\mu_0H_{c1}\approx4.0$~T, where N\'{e}el order disappears \cite{Povarov2019,Bhartiya2019}. Even though the spin Hamiltonian of \BaCd is, by now, very well characterized \cite{Bhartiya2021}, the origin of this high-field phase remains unclear. In particular, it persists in the magnetic field all the way up to $H_{c_2}=6.5$~T. Such a wide field range is difficult to reconcile even with predictions based on the simplistic magnon-pair mechanism \cite{Shannon2006,Ueda2013}. Most recent theoretical studies take magnon-pair interactions into account and conclude that the stability range for the spin-nematic phase must be lower by yet another order of magnitude \cite{Jiang2023}. Another concern is the subtle, but relevant, magnetic anisotropies in different $AA^\prime$VO(PO$_4$)$_2$ crystal structures, such as complex Dzyaloshinskii-Moriya (DM) interactions and patterns of $g$-tensor canting.

Here, we report the results of an NMR study aimed at better understanding the saturation process. We show that, unlike the presaturation phases of \PbV and \SrZn, the high-field phase in \BaCd has no dipolar order of any kind. At the same time, measurements of NMR relaxation rate independently confirm the high-field continuous phase transition at $H_{c2}$. Finally, we show that the in-between phase is characterized by very unusual spin relaxation that is dominated by two-magnon processes.
\vspace{-0.5cm}
\section{Material and Experimental Details}
\begin{figure}[t!]
	\centering
	\includegraphics[width=\columnwidth]{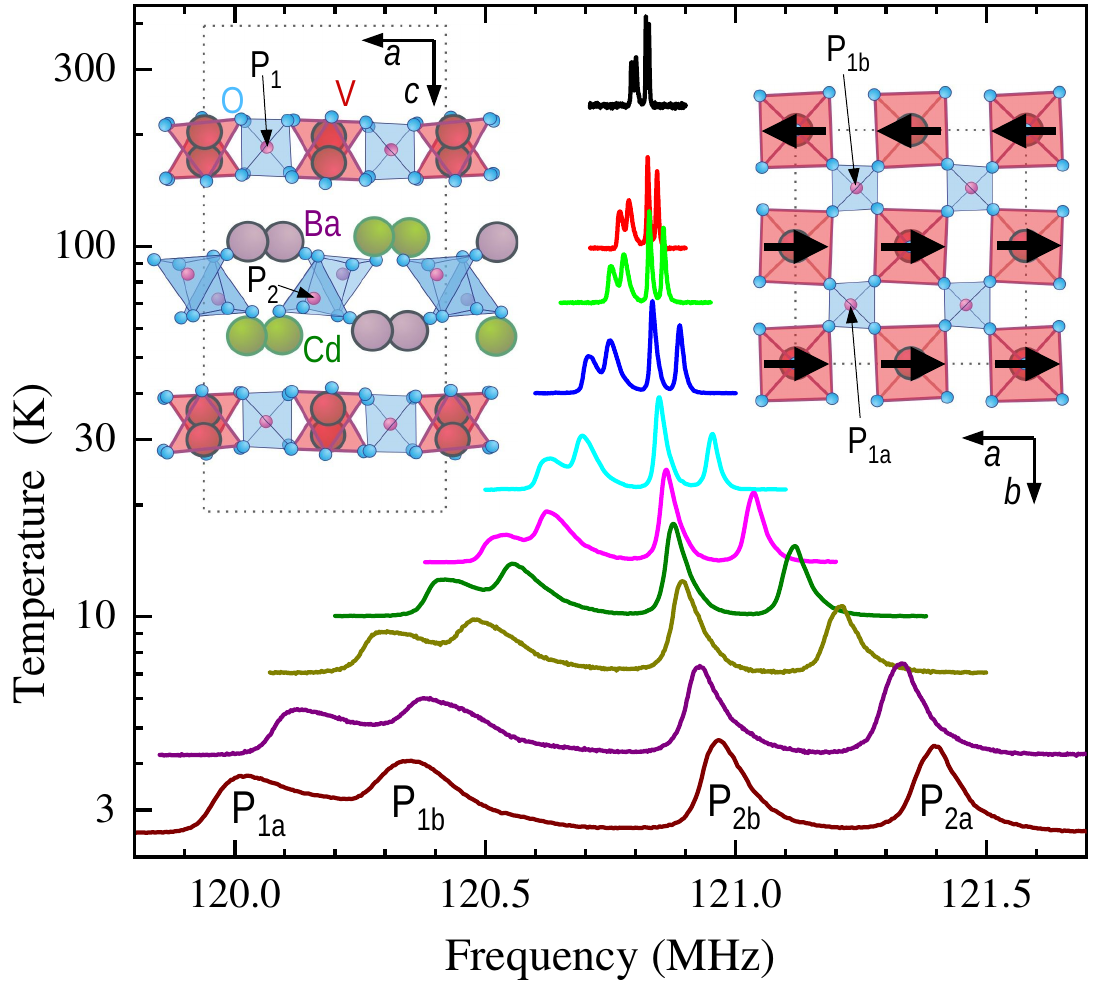}
	\caption{Temperature dependence of $^{31}$P NMR spectra of \BaCd in a field of 7~T applied parallel to the $c$ axis, from 235 to 2.6~K. Each spectrum is normalized to its maximum and vertically offset according to the temperature in the logarithmic scale. Insets show the crystallographic structure \cite{Bhartiya2021} and define the labeling of the P sites and NMR lines: on the left (right) is a side (top) view on the vanadium spin-1/2 planes. In the top view, thick arrows show the low-temperature up-up-down-down spin order (where only up-down-down spins are visible) \cite{Skoulatos2019} and the corresponding labeling of the two P$_1$ NMR lines.}
	\label{NewFigure}
\end{figure}

\vspace{-0.2cm}
The crystal structure (insets in Fig.~\ref{NewFigure}) and geometry of magnetic interactions in \BaCd are discussed in detail elsewhere \cite{Bhartiya2021}. Here, we recall only that  at room temperature the space group is orthorhombic, $P_{bca}$, with lattice parameters $a=8.84$ \AA, $b=8.92$ \AA, and $c=19.37$~\AA. There are $2 \times 4 = 8$ magnetic  $S=1/2$ V$^{4+}$ ions per cell, arranged in proximate square lattice layers in the $(a,b)$ plane, two layers per cell.
The symmetry is further lowered to $P_{ca}21$ in a structural transition at $T_s\approx 240$~K, resulting in eight distinct NN and NNN superexchange paths within each layer. Magnetic order sets in at  $T_N \approx 1.05$~K and is of an ``up-up-down-down" character (right inset in Fig.~\ref{NewFigure}) \cite{Skoulatos2019}, naturally following the alternation of NN interaction strengths along the crystallographic $b$ axis \cite{Bhartiya2021}.

NMR experiments were performed on a green transparent single-crystal sample ($3.2 \times 2.5 \times 0.35$~mm$^3$) from the same batch as those studied in Refs.~\cite{Povarov2019,Bhartiya2019,Bhartiya2021}. The sample was oriented to have the $c$ axis parallel to the applied magnetic field $\bm{H}$, and was placed inside the mixing chamber of a $^3$He-$^4$He dilution refrigerator for the measurements below 1~K and in the standard cold-bore variable-temperature insert (VTI) for temperatures $\geq 1.4$~K.

\begin{figure}[t!]
	\centering
	\includegraphics[width=\columnwidth]{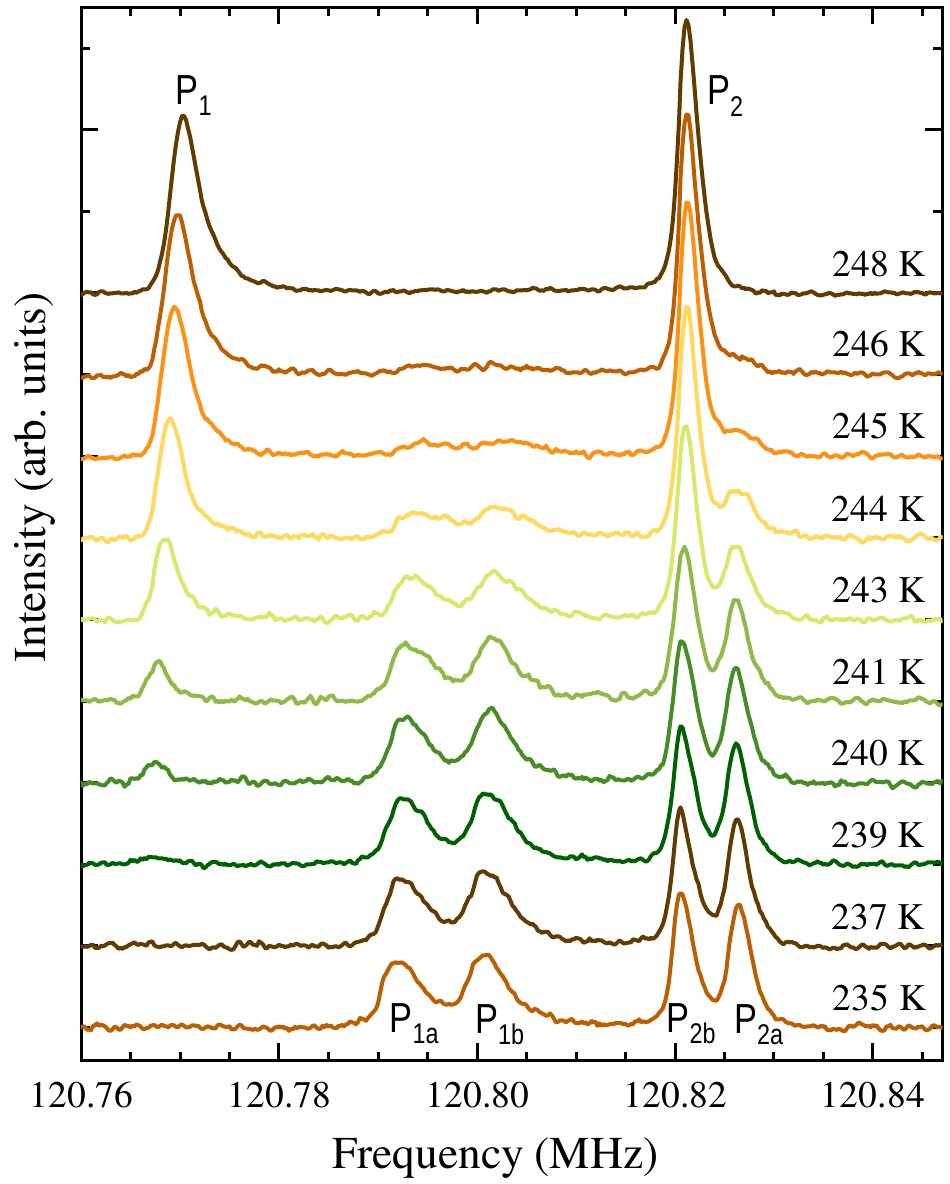}
	\caption{Temperature dependence of $^{31}$P NMR spectra of \BaCd in a field of 7~T applied parallel to the $c$ axis, focusing on the structural phase transition at $T_s\sim 240$~K.}
	\label{spectra_high_T}
\end{figure}

The $^{31}$P NMR spectra were taken by the standard spin-echo sequence and the frequency-sweep method. The nuclear spin-lattice relaxation rate $T_1^{-1}$ was measured by the saturation-recovery method, and the time $t$ recovery of the nuclear magnetization $M(t)$ after a saturation pulse was fitted by the stretched exponential function, $M(t)/M_0 =1-C\exp[-(t/T_1)^{\beta_s}]$, where $M_0$ is the equilibrium nuclear magnetization, $C \cong 1$ accounts for the imperfection of the excitation (saturation) pulse, and  $\beta_s$ is the stretching exponent to account for possible inhomogeneous distribution of $T_1^{-1}$  values \cite{Johnston2006,Mitrovic2008}. We find $\beta_s$ values close to 1 above 1~K, indicating a homogeneous system, and we find they decrease somewhat at low temperature, e.g., leading to values of 0.7$-$0.85 at 0.27~K in the 3.2$-$7.4~T magnetic field range, as shown in Fig.~\ref{relax_H}(b). Only one point in this figure has $\beta_s = 0.5$, corresponding to the obviously inhomogeneous phase mixture at the phase transition point.

\section{Experimental results}
\vspace{-0.2cm}
\subsection{NMR spectra}
\vspace{-0.2cm}
Figure~\ref{spectra_high_T} shows the evolution of the $^{31}$P NMR spectra across the previously mentioned structural phase transition. At 248~K we see two NMR lines corresponding to the \emph{two} phosphorus sites of the structure (left inset in Fig.~\ref{NewFigure}), as also observed by NMR in the paramagnetic and fully polarized phases of \PbV \cite{Nath2009, Landolt2020} and in \SrZn \cite{Ranjith2022} vanadates: The P$_1$ site is localized inside the planes of V$^{4+}$ spins; it is thus more strongly coupled to them, and corresponds to the broader NMR line at lower frequency. The P$_2$ site is localized between these planes; it is less coupled to the spins and corresponds to the narrower NMR line at higher frequency. Below the structural phase transition appears a modulation in the $b$-axis direction which \emph{doubles} the number of \emph{all} crystalline sites (see Supplemental Material of Ref. \cite{Bhartiya2019}). In particular, this creates an alternation of two different V$^{4+}$ spin sites (see Fig.~1 in Ref.~\cite{Bhartiya2021}) as well as an alternation of the two P$_{1a}$ and P$_{1b}$ sites in the planes of the spins and the two P$_{2a}$ and P$_{2b}$ sites in between these planes. In the spectra at 235 and 237~K we see the \emph{four} corresponding NMR lines: there is a low-frequency pair of broader P$_{1a}$ and P$_{1b}$ lines and a high-frequency pair of narrower P$_{2a}$ and P$_{2b}$ lines, where by index ``a'' (``b'') we label the outer (inner) lines. Finally, as expected for a first-order phase transition, the NMR spectra between 239 and 246~K present a mixture of the high- and low-temperature spectra, reflecting the mixed-phase region.

\begin{figure}
	\centering
	\includegraphics[width=\columnwidth]{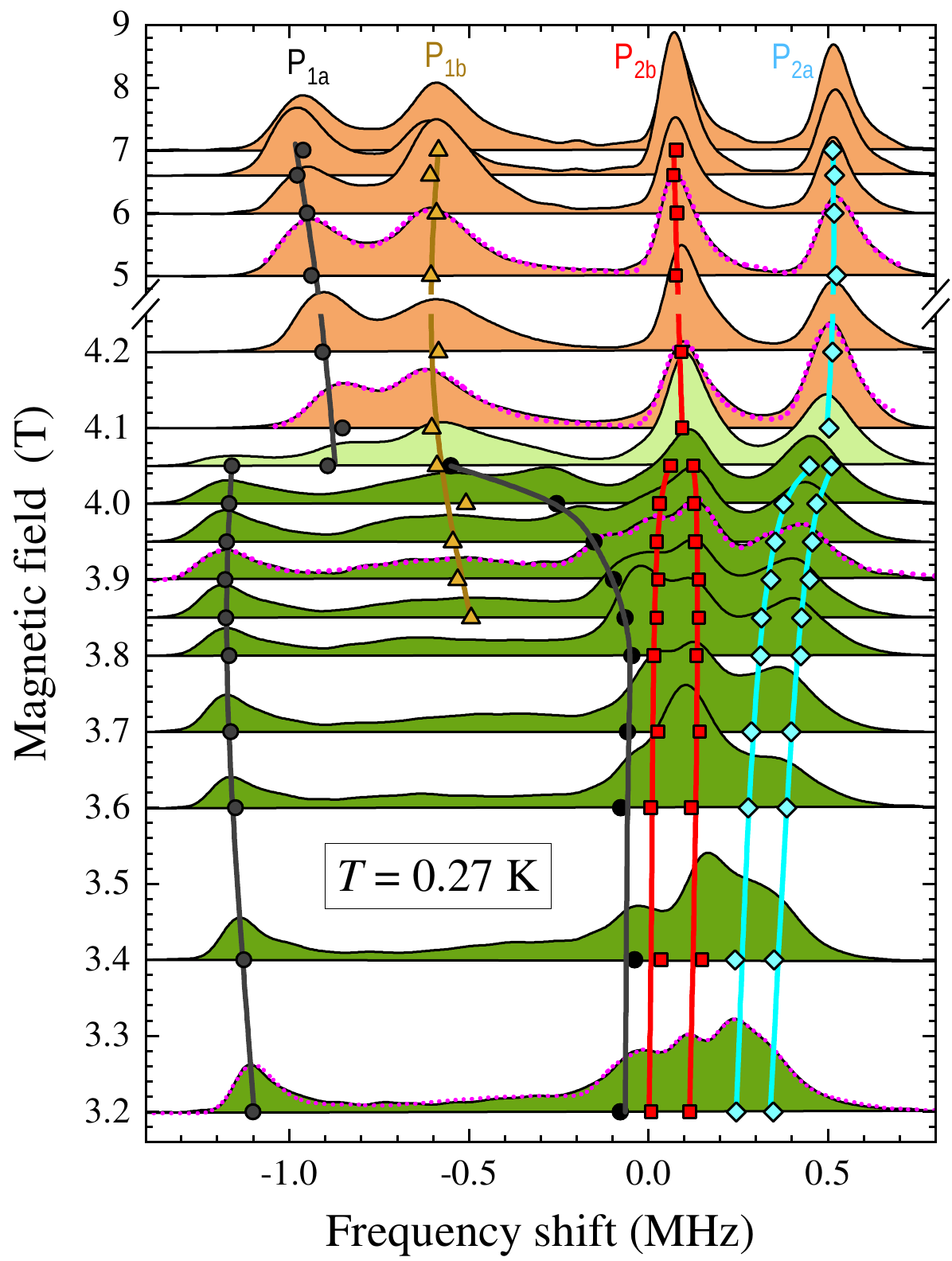}
	\caption{Magnetic field dependence of the low-temperature (0.27~K) $^{31}$P NMR spectra, plotted as a function of the frequency shift with respect to the Larmor frequency $^{31}\gamma \mu_0 H$ ($^{31}\gamma$~= 17.236~MHz/T) in order to reflect the local spin polarization value. Each spectrum is normalized to its integral and offset vertically according to the field value. Symbols are the peak positions obtained by fitting individual spectra, as described in the text. A few representative fits are shown by dotted magenta lines. The thick curves drawn through the symbols are guides for the eye.}
	\label{spectra_low_T}
\end{figure}

\begin{figure}
	\centering
	\includegraphics[width=\columnwidth]{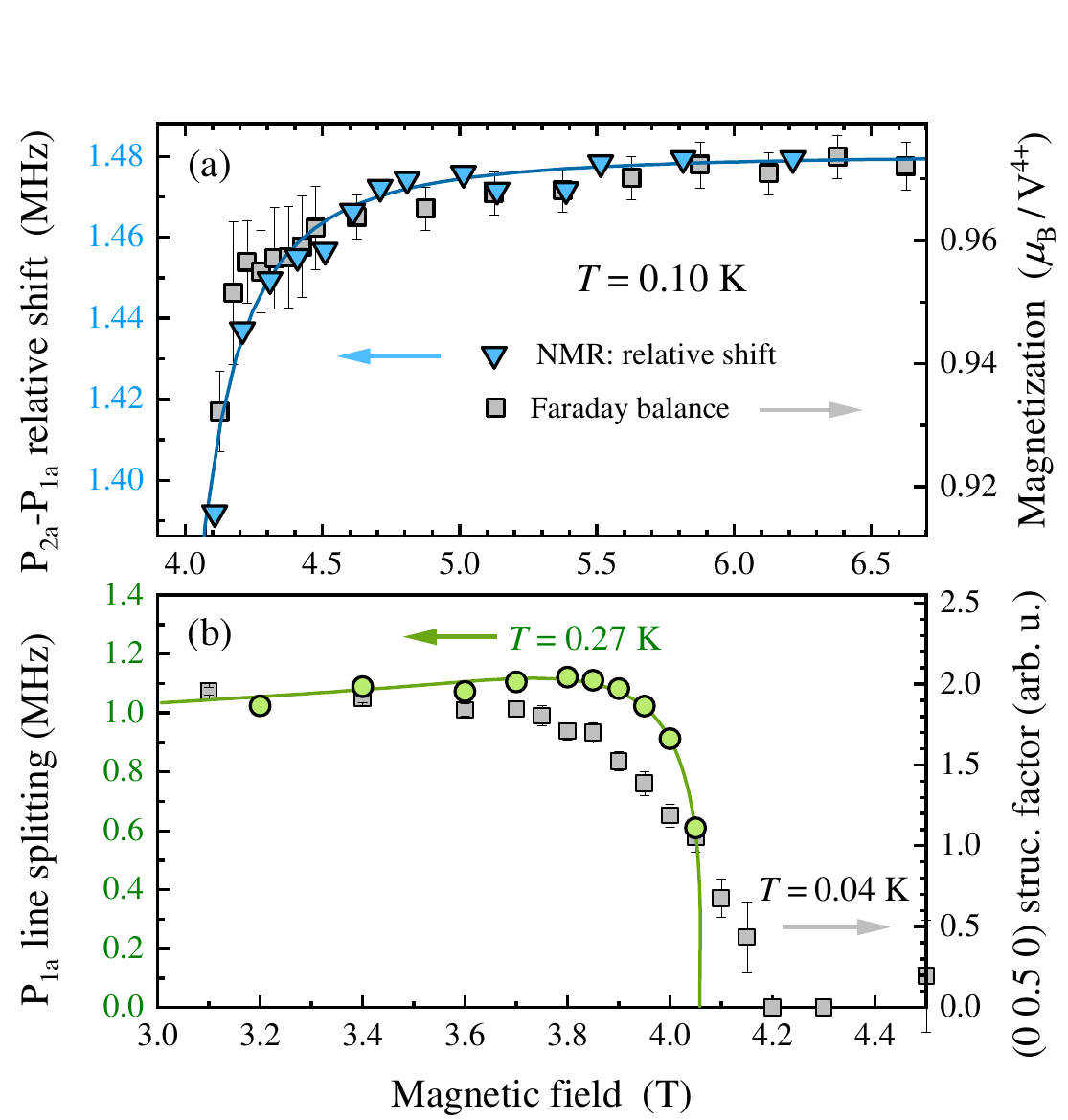}	
	\caption{(a) Field dependence of the sample magnetization above $H_{c1}$ at 0.10~K, deduced from the relative frequency shift of the two external NMR lines (triangles), in comparison to direct Faraday-balance magnetometry data (squares) from Ref.~\cite{Bhartiya2019} and the prediction (solid curve) given by Eq.~(\ref{simple}) (see the text). (b) Circles present the field dependence of the magnetic order parameter at 0.27~K, deduced from the line splitting of the P$_{1a}$ NMR line below $H_{c1}$ (Fig.~\ref{spectra_low_T}). The curve is a guide for the eye. Squares present the structure factor of the magnetic Bragg reflection $(0,1/2,0)$ at 0.04~K, obtained from the neutron data in Ref.~\cite{Bhartiya2019}.}
	\label{fit_results}
\end{figure}

As the temperature is lowered, the spectra taken at 7~T (Fig.~\ref{NewFigure}) retain the same shape, whereas their width grows proportionally to the magnetization \cite{Nath2008, Povarov2019} and approaches the low-temperature limit below 3~K: Compared to the spectrum at 2.6 K, the corresponding low-temperature spectrum measured at $T=0.27$~K is only 8\% wider, as shown in Fig.~\ref{spectra_low_T}, which presents the low-$T$ magnetic field dependence of the spectra. In Fig.~\ref{spectra_low_T} we see that the shape of the spectra does not change with the field down to 4.1~T, which is a clear signature that above $\mu_0H_{c1} \approx 4.05$~T the system is homogeneously polarized, without any dipolar magnetic order. A quantitative analysis of the spectra collected above $H_{c1}$  was obtained in a rather straightforward four-peak fit, where the slightly asymmetric shape of each NMR line was described by a Gaussian-based function that was \textit{ad hoc} modified by a damped third order term: $\propto \exp [-(\frac{x-x_c}{w})^2+b(\frac{x-x_c}{w})^3/(1.0+c|\frac{x-x_c}{w}|^3)]$. The  asymmetry-defining parameters were optimized on the 5~T spectrum to the $b = 0.30$ and $c = 0.0075$ values (fit shown in Fig.~\ref{spectra_low_T}) and then kept fixed to ensure the same line asymmetry for all fits. The thus obtained peak positions $x_c$ are shown in the figure by symbols. They carry local information on the spin polarization, that is, magnetization $M$: The latter stands in a linear relation with the frequency shift of each line. In order to get the most precise data, the temperature dependence of $M$ was monitored by following the separation between the highest-frequency (P$_{2a}$) and lowest-frequency (P$_{1a}$) lines, i.e., the relative frequency shift of the two lines that are most strongly coupled to the magnetization (see Fig.~\ref{NewFigure}). In adition, using \emph{relative} shift avoids the uncertainty of the magnetic field calibration. In order to better approach the zero-temperature values and thus exclude thermal effects \cite{Pregelj2020}, the line positions of these two NMR lines were remeasured at 0.10~K, and the field dependence of their relative shift for $H>H_{c1}$ is compared to the Faraday-balance magnetization data from Ref.~\cite{Bhartiya2019} in Fig.~\ref{fit_results}(a), where the two $y$ axes are scaled by 1.52~MHz/$\mu_B$ \cite{hyperfine}. The agreement is very good, proving the previous observation that $H_{c1}$ corresponds to a transition towards full saturation and that the magnetization process continues to higher fields.

Below $H_{c1}$ and at low temperature, a spontaneous AFM ordering occurs \cite{Skoulatos2019,Bhartiya2019}, so that each NMR line is expected to be split in two, where the difference in the coupling strength (the hyperfine coupling tensor) is expected to give much stronger splitting for the P$_1$ lines than for the P$_2$ lines, as was observed in \PbV \cite{Nath2009, Landolt2020}. In \BaCd (Fig.~\ref{spectra_low_T}) this creates quite complicated spectra with overlapping lines, from which we can reasonably recognize that the P$_{1a}$ line is indeed strongly split, P$_{1b}$ is unexpectedly only strongly broadened without visible splitting, and the two P$_2$ lines are weakly split as expected. In regard to the absence of AFM splitting of the P$_{1b}$ NMR line, it has a simple explanation in the up-up-down-down stripe type of AFM order \cite{Skoulatos2019} (right inset in Fig.~\ref{NewFigure}): P$_1$ phosphorus sites are approximately centered within 4 neighboring V$^{4+}$ spins, meaning that one type of P$_1$ sites ``sees'' either 4 up or 4 down spins while the other type always ``sees'' 2 up and 2 down spins. It is then obvious that the former sites observe strong AFM local field, providing a strong splitting of the P$_{1a}$ NMR line, while for the latter ones the local field approximately cancels out, leading to the absence of splitting of the P$_{1b}$ NMR line.

In order to distinguish the overlapping lines, we have selected the NMR spectrum at 3.9~T, which visibly provides the best line separation, enabling us to perform a reliable fit to 7 independent line positions (fit shown in Fig.~\ref{spectra_low_T}) and thus to fully define all three line splittings. As the splittings are all proportional to the Néel order parameter (OP) of the AFM phase, which is the staggered transverse spin component, their relative size (ratio) should be field independent. We used this additional constraint to stabilize the fits of all the other low-field spectra that are less well resolved: In these fits we fixed the two ratios of the AFM line splitting to the values obtained at 3.9~T. Thus obtained line positions are shown in Fig.~\ref{spectra_low_T} by symbols, and the corresponding splitting of the P$_{1a}$ line, providing an NMR image of the OP, is presented in Fig.~\ref{fit_results}(b) and compared with the corresponding information given by neutrons: the structure factor (square root of intensity) of the $(0,1/2,0)$  magnetic Bragg reflection reported in Ref.~\cite{Bhartiya2019}. One obvious discrepancy between the two data sets is that some residual neutron intensity persists above the transition. That is a known effect due to the finite wave vector and energy resolution of a neutron instrument, which is unable to differentiate between infinitely sharp Bragg peaks due to long-range order and broader scattering around the Bragg peak positions due to critical fluctuations.
It is more interesting that, unlike the neutron measurement, our NMR experiment indicates non-monotonic behavior of the order parameter. Here, we observe that both techniques are sensitive to a conceivable rotation of the OP (direction of canting), which is quite possible in the presence of complex DM interactions: NMR line splitting is affected by the angle dependence of the hyperfine coupling constant, and neutron intensities are affected by the so-called polarization factor. Nevertheless, the data shown in Fig.~\ref{fit_results}(b) rather point to a quite flat field dependence of the Néel OP, which is also what has been theoretically predicted for quantum spin-nematic candidates close to the nematic phase (see Fig.~5 in Ref. \cite{Jiang2023}).

\subsection{$T_1^{-1}$ relaxation rate}
\begin{figure}
	\centering
	\includegraphics[width=\columnwidth]{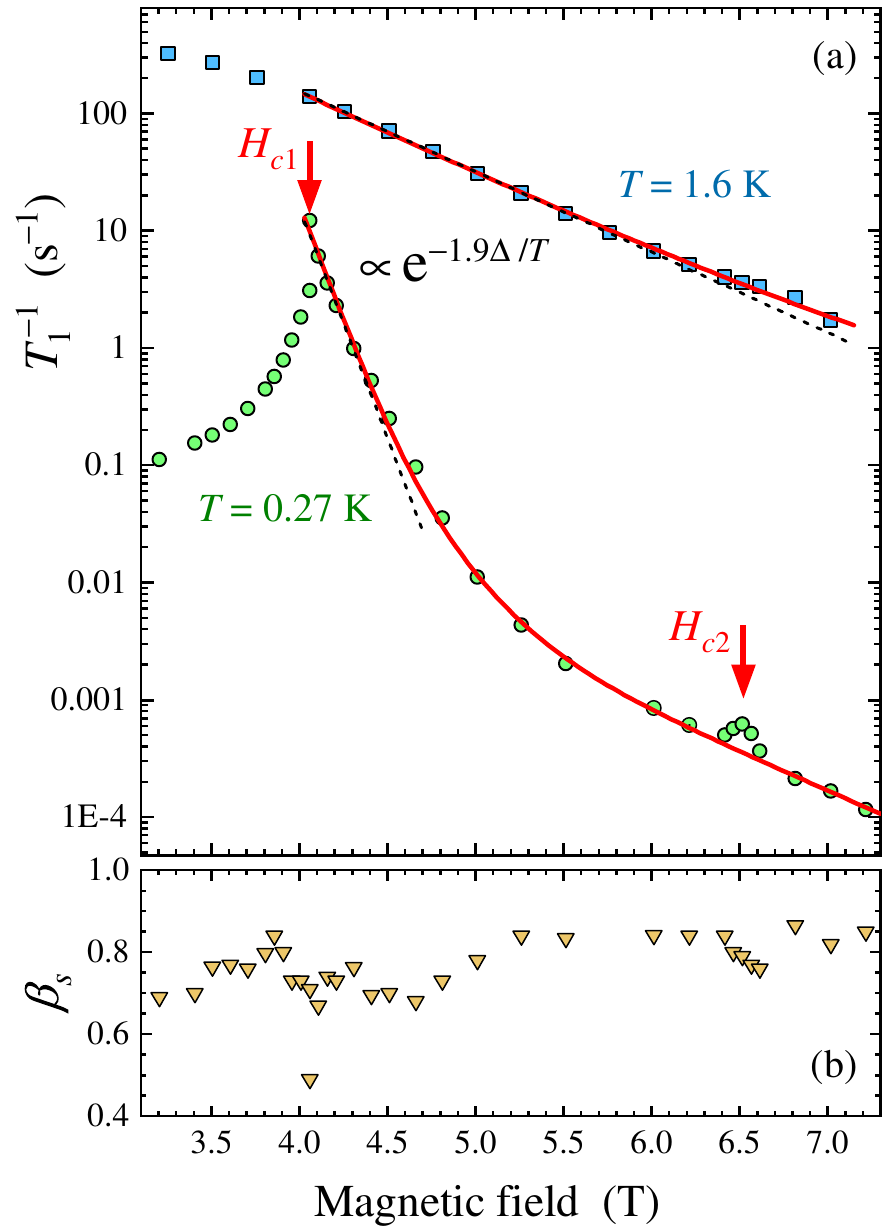}	
	\caption{(a) The $T_1^{-1}$ relaxation rate measured in \BaCd at 1.6~K (squares) and 0.27~K (circles) as a function of magnetic field applied along the $c$ axis. Experimental error bars are smaller than the symbol size. Straight dashed lines show activated behavior with energy gaps that scale with the single-magnon gap $\Delta(H)$, and red solid curves are the global three-exponential fit given by Eq.~(\ref{threeexp}), as explained in the text. (b) Field dependence of the stretching exponent $\beta_s$ from the fits to the measured relaxation curves at 0.27~K.}
	\label{relax_H}
\end{figure}

To better understand the low-energy spin dynamics below and above saturation, we performed measurements of the $T_1^{-1}$ relaxation rate vs field and temperature. The field dependence at two fixed temperatures is shown in Fig.~\ref{relax_H}(a) on a logarithmic scale. The most obvious features in the $T_1^{-1}(H)$ plot at $T=0.27$~K are the cusps corresponding to the phase transitions at $H_{c1}$ and at $\mu_0H_{c2} \approx 6.50$~T. While the latter appears weak on the logarithmic scale, it still corresponds to a twofold increase of the relaxation rate, a clear sign of a continuous phase transition. Its position exactly corresponds to the $H_{c2}$ transition previously detected as a $\lambda$ anomaly in calorimetry experiments \cite{Bhartiya2019}. The big $T_1^{-1}(H)$ peak obviously corresponds to the critical spin fluctuations related to the phase transition at $H_{c1}$ and tells us that this transition is essentially of the second order. However, in Fig.~\ref{spectra_low_T} we see that the spectrum taken at $H_{c1}$ corresponds to a \emph{mixture} of the two phases, and we could even measure the two different $T_1^{-1}$ rates corresponding to each of these two phases [Fig.~\ref{relax_H}(a)]. Similar behavior was also observed in \PbV \cite{Landolt2020} and in other quantum spin systems at the phase boundary of the low-temperature AFM phase, reflecting the weak first-order character of the transition, as predicted for a spin system that presents some magnetoelastic coupling \cite{Amore2009,Wulf2015}.

The observed behavior of the relaxation rate above $H_{c1}$ is superficially similar to that previously seen in \SrZn \cite{Ranjith2022}; however, a quantitative analysis  reveals substantial discrepancies. In all cases we consider relaxation via thermal activation across an energy gap that scales proportionately to the single-magnon gap $\Delta(H) = g\mu_B\mu_0(H-H_{c1})/k_B$ (here in kelvin units), where $g = 1.92$ is the $g$ factor \cite{Povarov2019} and $\mu_{\rm B}/k_{\rm B}$~= 0.67171~K/T. Furthermore, the log scale of Fig.~\ref{relax_H}(a) converts an exponential dependence $\propto e^{-\alpha \Delta/T}$ into its exponent, which depends linearly on field, and $\alpha$ is defined by the magnitude of the slope of the apparent linear dependence. While for a single-magnon condensation in the three-dimensional (3D) regime just above the saturation field this slope corresponds to $\alpha_0 \approx 3$ \cite{Ranjith2022}, from Fig.~\ref{relax_H}(a) we find that the initial value of the slope is very close to 2, which is expected \emph{above} the saturation field of a spin-nematic phase. In more detail, fitting the initial slope at 0.27~K in the 4.05--4.3~T range, we get $\alpha=1.90(9)$, and at 1.6~K for the 4.05--6.0~T range we get the identical value $\alpha=1.91(2)$. In order to cover by the fit all the 0.27~K data up to the highest field (excluding the peak centered at $H_{c2}$), we used the \emph{three}-activated-term fit:
\be
T_1^{-1}(H)=A(e^{-\alpha_0\Delta/T}+a_1 e^{-\Delta/T}+a_x e^{-\alpha_x\Delta/T}), \label{threeexp}
\ee
where the second term accounts for the expected presence of the \emph{single-magnon} relaxation and the third term additionally allows for an unexpected spin dynamics showing up at high field values (6.0--7.2~T). Using a global, \mbox{6-parameter} [$A(T_1),A(T_2),\alpha_0,a_1,a_x,\alpha_x$] fit over the \emph{both} data sets shown in Fig.~\ref{relax_H}(a) as a function of $\Delta(H) \propto H-H_{c1}\geq 0$, we indeed confirm that $\alpha_0=2.05(4)$ is equal to 2 within its experimental error. Equivalently, fixing $\alpha_0=2$ leads to a nearly indistinguishable fit, which is then preferred because it can be related to the two-magnon spin-nematic fluctuations. For this latter fit we find that the relative size of the single-magnon term amplitude (compared to the leading two-magnon term) is only $a_1=6.8(7)$\% and that the low-temperature high-field term is still much smaller, $a_x=0.15(3)$\%, and has a strongly reduced gap $\alpha_x=0.32(2)$.

\begin{figure}[t!]
	\centering
	\includegraphics[width=\columnwidth]{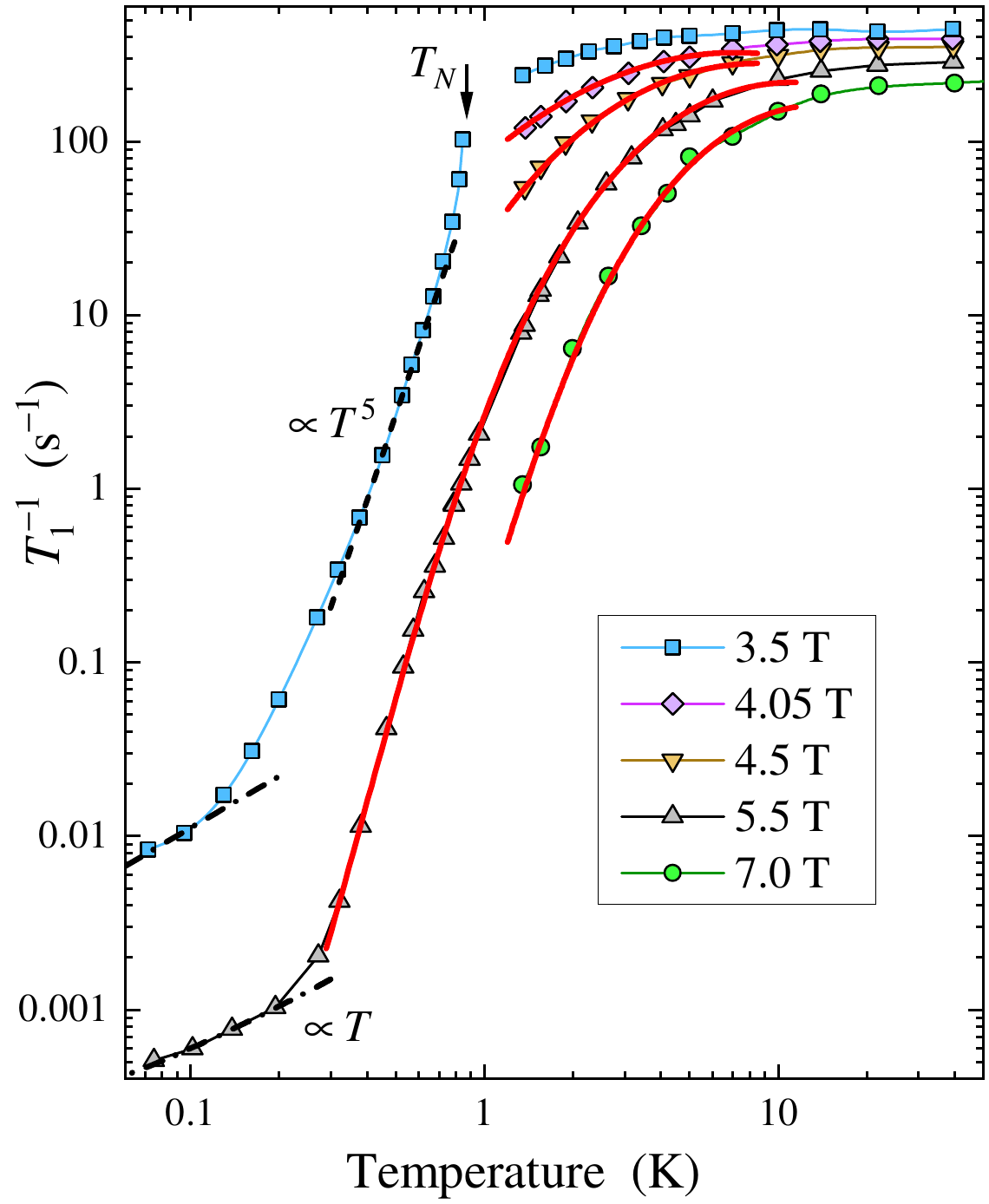}	
	\caption{The $T_1^{-1}$  relaxation rate measured in \BaCd as a function of temperature at several values of magnetic field applied along the $c$ axis (symbols).  Experimental error bars are smaller than the symbol size.  In a wide range above $H_{c1}$, the relaxation can be globally fit using the previously defined three-exponential fit [Eq.~(\ref{threeexp})], where the amplitude is taken to be a modified powerlaw $A \propto T^{\beta(T)}$, as explained in the text (red solid curves). At the lowest temperatures, $T_1^{-1}(T) \propto T$ (dash-dotted lines). In the N\'{e}el phase below $H_{c1}$, $T_1^{-1}$ has a characteristic $\propto T^5$ behavior (dashed line).}
	\label{relax_T}
\end{figure}

This analysis  suggests the overwhelming dominance of  the two-magnon processes. This is quite unusual and very different from what is seen in \SrZn, where it is the three- and single-magnon processes, expected for the single-magnon Bose-Einstein condensate (BEC) state, that provide the main relaxation mechanisms. Further confirmation for the prevalence of two-magnon relaxation in the high-field phase of \BaCd is found in temperature-dependent measurements at constant fields, as shown in Fig.~\ref{relax_T}. To analyze these data, we recall that, for a gapped system and purely parabolic $d$-dimensional magnon dispersion, the preexponential factor of the $T_1^{-1}$ rate is the power law $A(T) = A_0T^{\beta}$, where $\beta = d-1$. For the real dispersion relation of magnons in a specific material,  the exponent $\beta$ becomes temperature dependent, reflecting the ``effective" dimension of the system (see the Supplemental Material of Ref.~\cite{Mukhopadhyay2012}). We have thus fitted $T_1^{-1}(T)$ using the \emph{same} three-exponential function given by Eq.~(\ref{threeexp}) with the parameters defined in the previous fit, where  the common prefactor is taken to be an \textit{ad hoc} modified power law, $A(T) = A_0T^{\beta(T)}$, that allows for a suitable $\beta(T)=\beta_0-\beta_1\log(T/[\textrm{K}])$ dependence. This provided a remarkably precise \emph{global} fit of the data for all the field values $\geq H_{c1}$, covering as much as 1.5 orders of magnitude in temperature and nearly 5 orders of magnitude in $T_1^{-1}$ for the 5.5~T data in particular. The fit defines three parameters, $A_0 = 76(1)$~s$^{-1}$, $\beta_0 = 1.40(3)$, and $\beta_1 = 0.82(3)$, leading to the expected $\beta(T)$ dependence: at low temperature the system approaches the 3D regime, and $\beta$(0.3~K) = 1.8 is indeed close to the expected $\beta_{\textrm{3D}} = 2$ value, while the expected 2D regime value $\beta_{\textrm{2D}} = 1$ is reached at 3.1~K. The validity of the fit at its high temperature end is limited by the validity of the employed $A(T)$ dependence and the ``$T <$~gap'' condition for a description as an activated process.

The available low-temperature data show that at 5.5~T there is a continuity of the $T_1^{-1}(T)$ dependence down to the lowest temperature. This points to the continuity of the phase and the absence of any low-temperature ordering. There is only a crossover into an apparent linear dependence, observed below 0.2~K. A similar crossover seems to be observed below 0.1~K at 3.5~T, where the system is in the antiferromagnetically ordered phase, so such behavior may be associated with the Goldstone mode that is expected to show up at the low-temperature end if the axial symmetry is not broken \cite{Giamarchi1999}. However, considering the similarity with the 5.5~T data and the probable presence of the (subdominant) symmetry-breaking anisotropic terms of the Hamiltonian, such an interpretation is questionable. One might rather think of some common low-temperature physics that might be related to subleading terms of the system's Hamiltonian. Finally, inside the antiferromagnetically ordered phase, for the 3.5~T data in the intermediate temperature range (0.4--0.7~K), we find a strong power-law-like behavior, with the power exponent close to 5, which is usually the case in the BEC-type low-temperature phases of quantum antiferromagnets \cite{Mayaffre2000,Jeong2017,Orlova2018, Landolt2020}, reflecting a high-order relaxation process \cite{Beeman1968}.

\section{Discussion and conclusion}
In regard to the phase that appears above $\mu_0H_{c1}$~$\approx$ 4.05~T in \BaCd, our NMR results provide, in addition to a solid independent validation of the main findings and conjectures of Ref.~\cite{Bhartiya2019}, microscopic information that is crucial for defining the nature of these phases:

\emph{(i)} The low-temperature NMR spectra have exactly the same four-peak structure as those in the paramagnetic regime (Figs.~\ref{NewFigure} and \ref{spectra_high_T}) and are nearly field independent from $H_{c1}$ up to the full polarization saturation above 7~T (Fig.~\ref{spectra_low_T}). This provides clear evidence that the local spin polarization is homogeneous and lacks any transverse dipolar order, which is indeed one of the necessary (but not sufficient) characteristics of the spin-nematic phase \cite{Orlova2017}. The field dependence of the NMR line shift confirms and refines the previously observed field dependence of magnetization: The magnetization at $H_{c1}$ is slightly (by $\approx$~$-$6\%) reduced compared to full saturation and continues to increase in a wide range of fields above $H_{c1}$ [Fig.~\ref{fit_results}(a)]. With the measurements being carried out at 0.10~K, this dependence is certainly not of thermal origin \cite{Pregelj2020}, but it is not necessarily a consequence of some hidden order. It can also originate from a small anisotropy of the main exchange couplings. For an anisotropy of the $J_{xx} \neq J_{yy} \neq J_{zz}$ type, this has been discussed in the context of the LiCuVO$_4$ compound \cite{Orlova2017,HagemansCaux_PRB_2005_XYZchainsSaturation}. In the following paragraph we focus on the terms of the DM type, which are expected to provide the largest effect in \BaCd. We give the simplest estimate of this effect in the classical approximation and show that the observed field dependence is compatible with this mechanism [Fig.~\ref{fit_results}(a)]. Finally, we observe that recent theoretical predictions for the size of the magnetization variation in the spin-nematic phase \cite{Jiang2023} speak of a \emph{sizable} variation, definitely much bigger than the observed 6\%. Altogether, the static information is not really favorable to the existence of the spin-nematic phase in \BaCd.

The DM interactions can be rather complex in $AA^\prime$VO(PO$_4$)$_2$ structures \cite{Landolt2021}. To get a ballpark estimate, let us consider a toy model: We take the classical energy of two sublattices coupled by a single DM vector that is normal to the applied field. In the saturated state, DM coupling leads to a small canting of spins by an angle $\theta$ away from the field direction. This is driven by the gain in the DM energy per spin $E_{\textrm{DM}}(\theta)=-DS^2\sin(2\theta)/2$. It is stabilized by the corresponding loss of the Zeeman energy per spin $E_{Z}(\theta)=-gS\mu_B\mu_0H\cos \theta$, reduced by the modification of the exchange energy $E_{J}(\theta)=\tilde{J}S^2\cos(2\theta)/2$, where $\tilde{J}$ is some combination of exchange constants that will depend on how the two sublattices become canted. For small canting, the angle dependence of the total energy is quadratic, $E(\theta)=e_0-e_1\theta+e_2\theta^2$, and the equilibrium angle $\theta_0$ is defined by its minimum, $\theta_0=e_1/(2e_2)$, where $e_1=DS^2$ and $e_2=gS\mu_B\mu_0H/2-\tilde{J}S^2$. The depolarization ($1-\cos\theta_0$) is then:
\be
\frac{\delta M}{M} \approx \frac{1}{2} \left[\frac{DS}{g\mu_B\mu_0(H-\tilde{H})}\right]^2, \label{simple}
\ee
where $g\mu_B\mu_0\tilde{H}=2\tilde{J}S$.
Since the classical N\'{e}el state is the one that minimizes the exchange energy between its two sublattices, we can expect that $\tilde{H}<H_c$, with $H_c$ being the classical saturation field. The latter is equal to the single-magnon instability field and was previously shown to more or less coincide with $H_{c1}$ in \BaCd \cite{Bhartiya2021}. This simplest classical model can, indeed, reproduce the magnetization data in Fig.~\ref{fit_results}(a) (solid line) with the very reasonable values $\tilde{H}=3.75$~T and $DS^2=0.006$~meV. The latter corresponds to 1.6\% of the average $|J_1|$ and is comparable to the crude estimate based on the value of the spin flop field \cite{Bhartiya2019}: $D/J\sim(H_\text{SF}/H_{c1})^2\sim 1.5\%$.

\emph{(ii)} The most important contribution of NMR is certainly the insight into the low-energy spin fluctuations, measured by the $T_1^{-1}$ relaxation rate. First, the continuity of the $T_1^{-1}(H)$ dependence measured at 0.27~K definitely confirms the continuity of the phase between $H_{c1}$ and $H_{c2}$ [Fig.~\ref{relax_H}(a)]. The peak of $T_1^{-1}(H)$ centered at $H_{c2}$ is certainly a signature of some second-order phase transition, but the apparent continuity of the data below and above this peak indicates that the nature of the spin fluctuations is probably the same on both sides, suggesting that this might as well be the same phase, whereas the peak is signaling a phase transition appearing at lower temperature. This is strongly reminiscent of the observation of the low-temperature impurity-induced BEC phases in the doped \DTNX (DTNX) compound \cite{OrlovaBlinder2017,Dupont2017,DupontPRB2017,Orlova2018}, although in \BaCd it is not clear what the source of such impurities might be. Next, the continuity of the $T_1^{-1}(T)$ dependence down to 75~mK  measured at 5.5~T (Fig.~\ref{relax_T}) confirms the absence of any phase transition into an ordered state; only a crossover from an essentially activated to a low-temperature linear regime is observed at 0.2--0.3~K. This suggests that above $H_{c1}$ we see only a nearly fully polarized phase, presenting, at very low temperature, a crossover into some low-temperature physics, probably related to subdominant terms of the Hamiltonian and/or some defects and/or impurities.

The $T_1^{-1}$ data and fits shown in Figs.~\ref{relax_H}(a) and \ref{relax_T} provide clear and quantitatively precise identification of the largely dominant two-magnon relaxation process. Such a relaxation, expected \emph{above} the second critical (= saturation) field of the spin nematic phase, should be regarded as one of the key signatures of this phase. This type of relaxation was already observed above the saturation field of volborthite \cite{Yoshida2017}, below which a putative spin-nematic phase was clearly detected \cite {KohamaIshikawa_PNAS_2019_VolborthiteNematic}, but experimental precision of these NMR data was not sufficient to distinguish between the two- and  the three-magnon process. In any case, the two-magnon relaxation is \emph{not} expected above the \emph{first} critical field of such a phase, which would be the case if such a phase existed in \BaCd between $H_{c1}$ and $H_{c2}$.

Finally, recent theoretical simulations of a spin-nematic phase \cite{Jiang2023} point to a \emph{very narrow} field range where such a phase is expected, which is clearly not the case for the two critical fields of \BaCd, where $H_{c2} = 1.60 H_{c1}$. Furthermore, as already pointed out, $H_{c1}$ coincides with the theoretical prediction for a \emph{single}-magnon instability field \cite{Bhartiya2021}. Altogether, all the evidence points to the fact that $H_{c1}$ is, in fact, the saturation field of the system, whereas what is observed at $H_{c2}$ is related to some very low temperature ordered phase of undefined origin unrelated to the dominant terms of the system's Hamiltonian. The observed two-magnon relaxation process above $H_{c1}$ then implies that the system is very close to the spin-nematic instability, so that the corresponding fluctuations dominate the spin dynamics, but the spin-nematic phase is, in fact, \emph{not} stabilized in \BaCd. This situation appears to be archetypal for other spin-nematic candidate compounds, and our results establish NMR-based criteria to define the true nature of their high-field phases.

\begin{acknowledgments}
We are grateful to N. Shannon, T. Momoi and M. \mbox{Zhitomirsky} for stimulating discussions and to R. Nath for providing the original magnetic susceptibility data.
This work was supported, in part,  by the Swiss National Science Foundation, Division II.
\end{acknowledgments}

\bibliography{azbib_new2}

\end{document}